\documentclass[fleqn,usenatbib]{mnras}

\usepackage[T1]{fontenc}
\usepackage{ae,aecompl}
\usepackage[dvipsnames]{xcolor}
\usepackage{graphicx}	% Including figure files
\usepackage{amsmath}	% Advanced maths commands
\usepackage{amssymb}	% Extra maths symbols
\usepackage{listings}
\usepackage{soul,xcolor}

\newcommand{\msun}{M$_\odot$}
\setstcolor{red}

\title[PBH simulations]{Dark Matter Simulations with  Primordial Black Holes in the Early Universe}

\author[M. V. Tkachev, S. V. Pilipenko,  G. Yepes]{
Maxim V. Tkachev$^{1}$\thanks{mtkachev@asc.rssi.ru (MVT)}
Sergey V. Pilipenko$^{1}$\thanks{spilipenko@asc.rssi.ru (SVP)},
and Gustavo Yepes$^{2,3}$
\newauthor
\\
$^{1}$Astro Space Center, P. N. Lebedev Physical Institute of RAS, Profsojuznaya 84/32, Moscow 117997, Russia\\
$^{2}$Departamento de F\'{i}sica Te\'{o}rica, M\'{o}dulo 8  Universidad Aut\'{o}noma de Madrid, 28049 Madrid, Spain\\
$^{3}$Centro de Investigaci\'{o}n Avanzada en F\'{i}sica Fundamental (CIAFF), Universidad Aut\'{o}noma de Madrid, 28049 Madrid, Spain\\
}

\date{Accepted XXX. Received YYY; in original form ZZZ}

\pubyear{2020}

\begin{document}
\label{firstpage}
\pagerange{\pageref{firstpage}--\pageref{lastpage}}
\maketitle

% Abstract of the paper
\begin{abstract}
Primordial Black Holes (PBH)  with masses of order $10-30 M_\odot$ have been  proposed as a possible explanation  of the gravitational waves emission events recently  discovered by the  LIGO observatory. If true, then  PBHs would constitute a  sizeable fraction of the dark matter component in the Universe. Using a series of cosmological N-body simulations which include both dark matter and a variable fraction of  PBHs ranging from  $f_{PBH} = 10^{-4}$ to $f_{PBH} = 1$,  we  analyse the processes of formation and disruption of gravitationally bound PBH pairs, as well as the merging of both bound and unbound pairs, and estimate the probabilities of such events.  We show that they are in    good agreement with the constrains to the PBH abundance  obtained by the LIGO and other research groups. We  find that pair stability, while being a main factor responsible for the merger rate, is significantly affected by the effects of dark matter halo  formation and clustering. 
As a side result,  we also  evaluate the effects of numerical errors  in the stability of  bound pairs, which can be useful for  future research using this methodology. 

\end{abstract}

\begin{keywords}
cosmology-simulations --- dark matter --- primordial black holes
\end{keywords}

%%%%%%%%%%%%%%%%% BODY OF PAPER %%%%%%%%%%%%%%%%%%

\section{Introduction}  \label{sec:intro}
Various astrophysical and cosmological observations provide substantial evidences that  firmly establish  the existence of a dominated  collisionless,  non-baryonic,  dark matter (DM) component  in our Universe. However, the nature of DM remains one of the major unsolved problems in fundamental physics.

Primordial black holes (PBHs),  originally proposed for the first time by  
\citet{zel_nov},  can be formed during the  early stages of the evolution of the Universe due to the  collapse of large energy density fluctuations. They have been proposed as promising candidates for DM \citep{dm1, dm2, dm3, dm4, dm5, dm6, dm7}.  PBHs have recently received quite a lot of attention  in relation to the events observed by The Advanced Laser Interferometer Gravitational-Wave Observatory (LIGO) \citep{ligo1, ligo2, ligo7, ligo3, ligo5, ligo4, ligo8}\footnote{LIGO/Virgo Public Alerts from the O3/2019 observational run can be found at https://gracedb.ligo.org/superevents/public/O3/}.

Although, at present,  the hypothesis of PBH existence is yet neither proven nor refuted, the very observational limits on its abundance represent themselves a powerful and unique method of investigating the early Universe at small scales, which cannot be tested by any other method \citep{pbh_obs1, pbh_obs2, pbh_obs3}.
The most important potential bounds on the PBH abundance in the mass range around $10-30$ \msun  can be obtained from LIGO observations, assuming, that they involve the  merging of  primordial black holes pairs. Despite the fact, that some of the LIGO events \citep{ligo6} are  now accepted as having astrophysical origin \citep{astro_bh}, the origin of some other events might appear to be primordial \citep{ligo8}.  If PBHs made up a significant fraction of DM, the PBH merger rate  would produce a  gravitational wave (GW) background which would be much larger than what is currently observed by LIGO \citep{constr1, constr2, yacine}, as well as other effects, such as gravitational lensing of supernovae \citep{sn_lim}. However,  SN lensing constraints were reanalysed by \citet{fleury} and showed  to agree with observations. Likewise, LIGO constraints are also  subject of  large theoretical uncertainties.

A reliable estimate of the GW signatures from PBH binary mergers requires a good understanding of the process of formation of  PBH binaries and their subsequent interactions with the surrounding matter that may disrupt the binaries (for example, see \citet{raidal} or \citet{trashorras}). Despite the fact that the bulk of this work is dedicated to  PBH binaries, we also consider the effects of hyperbolic\footnote{In this context, "pair" is not necessarily a \textit{gravitationally bound} pair, unless specifically mentioned, and in general should be perceived simply as "two interacting particles".} close encounters (as suggested by \citet{nesseris}), and how a proper account of these events could affect our final results.

In this paper we present  a numerical  model  describing the evolution  of PBHs  with masses $\sim$ 30 \msun  throughout the earliest stages of the Universe and up  to  redshifts $z \approx 3.0$. Despite the aforementioned constraints on the abundance of PBHs of such masses, in our model, just  for  testing purposes,  we also assume  the extreme case where  all of the DM consists of PBHs, i.e., $f_{PBH} = \frac{\Omega_{PBH,0}}{\Omega_{m,0}} = 1.0$, where $\Omega_{PBH,0}$ is the   density parameter of  PBHs at redshift $z = 0$, and $\Omega_{m,0}$ is the total matter  density parameter  at the same redshift. However, our main conclusions  are obtained from simulations with much smaller fractions, down  to $f_{PBH} = 10^{-4}$.

We run a series of dark-matter only cosmological N-body simulation, using a modified version of the TREEPM cosmological code \texttt{GADGET-2}. To our knowledge, previous to  our  work,  cosmological N-body codes have been    applied for studying PBH problems  only by  \citet{inman}). Although, they  pursued different goals, such as the detailed and wide-scale study of the interaction of "normal" and "PBH dark matter", while, as the  authors themselves mentioned in that paper, "their numerical setup does not have high enough spatial resolution to resolve individual binary orbits".
In our work, on the other hand, when $f_{PBH} < 1.0$ we split the  DM fluid  into \textit{"PBH dark matter"} represented as individual particles and a \textit{"normal dark matter"},  that is represented as a unperturbed, uniform medium. 
We then study the  complex processes of formation, disruption and merging of PBH pairs  from the orbits of the PBH particles  in the  context of a 2-body problem, where one body is a given PBH particle and the second body is one of its neighbours with the minimal total energy $E$.
The total energy is computed as a product of specific orbital energy $\epsilon$ and reduced mass $\mu = \frac{m_1 m_2}{m_1+m_2}$, where $m_1 = m_2 = M$ is is a mass of a PBH:
\begin{equation} \label{eq0}
E = \mu\epsilon = \frac{M}{2} \left( \frac{v^2}{2} - \frac{2GM}{r} \right),
\end{equation}
where $v$ is a relative velocities of the particles (including the Hubble velocity), $r$ is the radius vector in physical coordinates.
We evaluate the stability of a pair in general and its dependence on the numerical errors  associated to  gravitational softening parameter and timestep (see Section \ref{sec:analysis}). Additionally, as mentioned above, we also consider the  possibility of binary capture due to the general relativity (GR) effects from close hyperbolic encounters.
The merging process of both bound and unbound pairs is evaluated by estimating their rate of energy loss due to the radiation of gravitational waves. The merging condition is defined as
\begin{equation} \label{eq1}
\sum_{i=1}^{n} T\dot{|E|} \geq \frac{GM^2}{2|a|}
\end{equation}
where $\dot{E}$ is the rate of energy loss by the pair due to the radiation of gravitational waves, $T$ is the current orbital  period of the pair, and $a$ is a semi-major axis of the pair ($a<0$ for hyperbolic orbit). Since the majority of our pairs have eccentricities close to $e \approx 1$, we are safe to assume  that a pair radiates gravitational waves only when its particles pass the pericentre. Therefore, the  sum  is limited to the number of full periods,  $n$, during which a pair remained gravitationally bound, or 1 in case of hyperbolic orbit.
 The energy loss can be estimated as
\begin{equation} \label{eq3}
\dot{E} = \frac{64G^4 M^5}{5 c^5 a^5} g(e)
\end{equation} 
with
\begin{equation} \label{eq4}
g(e) = \begin{cases}\frac{24\cos^{-1}(-\frac{1}{e}) \left(1 + \frac{73}{24}e^2 + \frac{37}{96}e^4\right) + (e^2-1)^{1/2}(\frac{301}{6} + \frac{673}{12}e^2) } {(e^2-1)^{7/2}}, & e \geq 1, \\\\
\frac{24\pi}{(1-e^2)^{7/2}}\left(1 + \frac{73}{24}e^2 + \frac{37}{96}e^4\right ), & e < 1, \end{cases}
\end{equation}

and $e$ being the eccentricity of the pair, $G$ is the  gravitational constant, and $c$ is the speed of light in vacuum, respectively \citep{radiation1, radiation2}.
Both $E$ and $\dot{E}$ are calculated with orbital parameters that the pair has at a given moment. If the accumulated radiated energy of a given pair gets larger than $E$, and the pair is not disrupted during  this time, the pair is considered to be merged.
For an accurate description of the algorithm see Section \ref{sec:merger}.

The paper is organised as follows:
\begin{enumerate}
\item In Section \ref{sec:simulations} we describe the details of our numerical model and provide the list of parameters   used to run   the N-body simulations.
\item In Section \ref{sec:analysis} we calculate the probabilities of the formation of PBH pairs, starting during the radiation dominated epoch. Then,  we evaluate the possibilities for the pair to be disrupted due to numerical errors rather than  to physical processes.
\item In Section \ref{sec:merger} we provide our estimates of the merger rate of PBH pairs, depending on the fractional abundance  of PBH, including hyperbolic captures due to GR effects.
\item Finally,  we leave Section \ref{sec:conclusions}  for discussion regarding the question whether or not PBHs with masses of order of tens of solar masses can constitute a large enough fraction of dark matter to explain gravitational waves events as   merging of PBHs.
\end{enumerate} 

%%%%%%%%%%%%%%%%%%%%%
%%%%%%%%%%%%%%%%%%%%%
\section{N-body Simulations with Primordial Black Holes}  \label{sec:simulations}
We are mainly  interested  in finding and analysing the behaviour of PBH \textit{pairs}. Therefore, their orbits should be calculated with sufficient precision. Normally this would require the use  of  direct N-body methods, which are computationally very expensive for large number of particles.   We use instead  the publicly available N-body code   \texttt{GADGET-2}\footnote{http://wwwmpa.mpa-garching.mpg.de/~volker/gadget/} \citep{gadget}, which is widely used  for cosmological simulations. This code   uses a combined Tree + Particle Mesh (TreeMP)   algorithm to estimate the gravitational accelerations for each particle  by decomposing the gravitational forces into a long range term, computed from Particle-Mesh  methods, and  short scale interactions from  nearest neighbours using Tree methods.   The code is MPI parallel and  can be used  with periodic boundary conditions in the comoving frame.  The scaling of the code with the number of particles   is  $\cal{O}(N\log N)$ so  it can handle a large number of particles with reasonable computational resources.

The simulations were run in the 
\texttt{MareNostrum} supercomputer  at BSC-CNS \footnote{https://www.bsc.es/} using a modified version of  \texttt{GADGET-2} in which  the free parameters of the code  were chosen  such that  the  force computation was  accurate enough for PBH binaries to form and survive without being disrupted by pure numerical errors.  We took  the gravitational softening  length  $\epsilon = 0.001$ pc and a timestep of $\Delta t = 10^{-6}$ in $\log(a)$.  

These parameters  have a significant impact in the orbit stability, so we ensured that our choices are optimal and that a further reduction of both $\epsilon$ and $\Delta t$ does not yield any significant improvement in orbit stability.

Previous analytic estimates suggest  that the first   PBH \textit{pairs} begin to appear  approximately at the   radiation-matter energy equipartition  epoch ($z \sim 10^3$) \citep{nakamura, sasaki} or, depending on the PBH fraction, even earlier -- during the radiation dominated epoch. 
Therefore, we had to modify the  cosmological equations in the \texttt{GADGET-2} code  to  be valid during   the radiation dominated epoch  of the Universe, since we start our simulations at  $z_{init}=10^5$.

As was discussed earlier, in this work we focus  on   the formation of PBH pairs  for different fractions of PBH to  the total DM component. So, when  $f_{PBH} < 1.0$, we have to split  the DM into two types: \textit{"PBH DM"} and \textit{"normal DM"} (see Sec. \ref{sec:intro}).

As it is shown in Sec. \ref{sec:analysis}, in our simulations virialised structures start to form  as early as $z \approx 10^3$, which could be caused by the initial Poisson distribution of PBHs. This differs significantly from the case of standard $\Lambda$CDM model, where the  first virialised structures start to form at $z\simeq 40$. Therefore we  assume that  the   \textit{"normal DM"}  behaves as an  uniform medium in our simulations. Nevertheless, we will discuss later why the formation of virialised structures is important and how it affects the behaviour of PBH pairs (see Sec. \ref{sec:virial}).

\begin{figure}
    \centering
    \includegraphics[width=\linewidth]{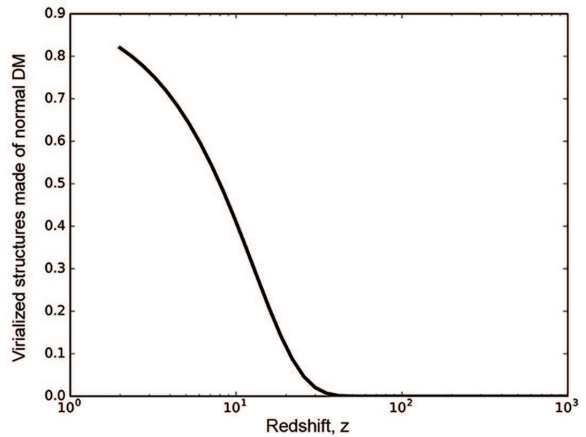}
    \caption{The fraction of virialised structures made of \textit{normal DM} for halos more massive than 30~M$_\odot$, as a function of redshift $z$
    }
    \label{fig:press}
\end{figure}

In order to quantify the range of redshifts  in which this approximation  is applicable, we calculate the fraction of virialised \textit{normal DM} using  the analytical  Press-Schechter formalism for halos more massive than 30~M$_\odot$ (see \citet{press}).
We find  that the fraction of virialised structures made of \textit{normal DM} is less than 50\% at $z>10$.
(see Figure \ref{fig:press}).

Thus, the presence of \textit{normal DM} influences the evolution of \textit{PBH DM} only through its contribution to the computation of  the Hubble function  in  the \texttt{GADGET-2} code:
\begin{equation}
H(z) = H_0\left [\Omega_{\Lambda,0}+\Omega_{m,0}(1+z)^3-\Omega_{r,0}(1+z)^4\right]^{\frac{1}{2}},
\end{equation}
For our  purposes,  this approach seems reasonable in terms of required  accuracy  and  reasonable  computational resources since we avoid  the costly calculations  of the interactions between  DM and PBH particles.

We have run a series of six  dark matter only simulations. Five of them   correspond to different  values of  the PBH fraction,   ranging from $f_{PBH} = 1.0$ to $f_{PBH} = 10^{-4}$.  For convergence study, we have also run an additional  simulation   with  the PBH fraction $f_{PBH} = 10^{-4}$  but increasing the number of particles. In  all  our simulations the PBH fraction is represented by up to  $N_{total} = 10^5$  particles with masses 30 \msun. The   initial conditions for PBH particles were set  at  $z = 10^5$. Their  coordinates   were  set  from the uniform distribution and  their initial peculiar velocities are equal to zero.  The high resolution simulation has  10 times more particles $N_{total} = 10^6$.   As   $f_{PBH} $ varies  from one simulation to another,  we adjust the box size accordingly in order to achieve the desired particle density. Therefore, the box size varies from $15.21$ $h^{-1}$~kpc for $f_{PBH} = 1.0$, to $ 706.04 \;h^{-1}$~kpc for $f_{PBH} = 10^{-4}$ (1521.12 $h^{-1}$~kpc for the high resolution simulation). 

In order to reduce possible  artefacts due to periodicity of the relatively small box, the final  redshift of the simulations was set to $z = 3$. A total of 200 snapshots were  stored  at redshift intervals equally-spaced in logarithmic scale, starting from $z = 10^5$ to $z = 3$.

All simulations  share  the same cosmological parameters  in agreement  with  the values obtained by  the \citet{planck},
i.e. $\Omega_{\Lambda,0}=0.692885$, $h=0.6777$, $\sigma_8=0.8288$, $n_s=0.9611$. The matter density is taken as $\Omega_{m,0}=0.307023$, while radiation density is taken as $\Omega_{r,0} = 0.000092$, where $\Omega_{m,0} + \Omega_{r,0} = 0.307115$ and, accordingly, $\Omega_{m,0} + \Omega_{r,0} + \Omega_{\Lambda,0} = 1.0$. 

A summary of the most relevant parameters of  the  simulations is given  in Table \ref{tab:sim}.

\begin{table*}
\caption{Most relevant parameters of the simulation suite.} 
\centering
\begin{tabular}{llllllll}
\hline
Fraction $f_{PBH}$        & $1.0$  & $10^{-1}$    & $10^{-2}$    & $10^{-3}$    & $10^{-4}$    & $10^{-4}$     \\ \hline
Box Size $(kpc/h)$         & 15.21 & 32.77 & 152.11 & 327.71 & 706.04 & 1521.12 \\
N. particles $N_{total}$     & $10^4$   & $10^4$    & $10^5$    & $10^5$    & $10^5$    & $10^6$     \\
Mass res. $(M_\odot)$       & 30   & 30    & 30    & 30    & 30    & 30     \\
Grav. soft. $\epsilon (pc)$       & $10^{-3}$    & $10^{-3}$     & $10^{-3}$     & $10^{-3}$     & $10^{-3}$     & $10^{-3}$      \\
Min. timestep $\Delta t_{min}$ & $10^{-6}$    & $10^{-6}$     & $10^{-6}$     & $10^{-6}$     & $10^{-6}$      & $10^{-6}$     \\
Initial redshift & $10^5$   & $10^5$    & $10^5$    & $10^5$    & $10^5$    & $10^5$     \\ \hline
\end{tabular}
\label{tab:sim}
\end{table*}

%%%%%%%%%%%%%%%%%%%%%
%%%%%%%%%%%%%%%%%%%%%
\section{Formation and stability of PBH pairs}   \label{sec:analysis}
In this Section we estimate  the PBH pair formation rate  and evaluate the possibilities for the pair to be disrupted due to numerical errors rather than to  physical processes.

For the analysis of the formation of pairs in our simulations, we calculate the total energy $E$ of a two body system for every particle and its closest neighbour (see Equation \ref{eq0}).
We consider a pair of particles to be gravitationally bound to each other if the  condition $E<0$ is fulfilled. Naturally, there are certain amount of pairs, which have more than one neighbour that is gravitationally bound to it. This amount varies from 1\% to 20\% of  the total \textit{pairs}, depending on redshift and  the fraction of PBH .  In the context of the method, which we use for the evaluation of the merger rate (see Section \ref{sec:intro}), as well as for simplicity, we still treat such pairs as \textit{binaries}, and only account for the neighbour with the minimal total energy value (which could not  necessarily be  the closest neighbour).

As we mentioned in \S \ref{sec:intro}, aside from classical Newtonian captures we also considered the  possibility that energy radiated as GW during close hyperbolic encounters might be sufficient to result in binding the given pair, i.e., $E-T\dot{|E|}< 0$. We found out that even though such events are possible, they are extremely rare and are not  present at all  for simulations with  $f_{PBH}<0.1$.   We compared the number of such events, $n_{cap(GR)}$ (i.e. GR-related capture) with the number of pure Newtonian captures, $n_{cap(Nt)}$, to gain some perspective. In  Table \ref{tab:GR} we quote the ratio of these two number as a function of  the PBH fraction of the different simulations.
Therefore, we are confident  that,  at least for our simulation set,  the proper accounting of  GR corrections to coordinates and velocities is probably not necessary. However, as we already mentioned, this effect is only present in simulations with higher values of $f_{PBH}$. But at the same time, these two simulations in our set have the least number of particles (N = 10000). Therefore, if we considered a larger simulation with similar values of $f_{PBH}$ ($N=10^6$, for example), this effect ought to be studied in more detail.

\begin{table}
\caption{The ratio  of the number of GR captures ($n_{cap(GR)}$) to the total number of pure Newtonian captures ($n_{cap(Nt)}$), for the simulations with different  PBH fraction ($f_{PBH}$)  } 
\begin{tabular}{ll}
$f_{PBH}$    & $n_{cap(GR)} / n_{cap(Nt)}$ \\
\hline
$1.0$ & 0.00011 \\
$10^{-1}$  & 0.00042  \\
$10^{-2}$  & 0.  \\
$10^{-3}$  & 0.  \\
$10^{-4}$  & 0.  \\
$10^{-4}(N_{total}=10^6)$ & 0.  \\
\hline
\end{tabular}
\label{tab:GR}
\end{table}

\begin{figure}
    \includegraphics[width=0.47\textwidth]{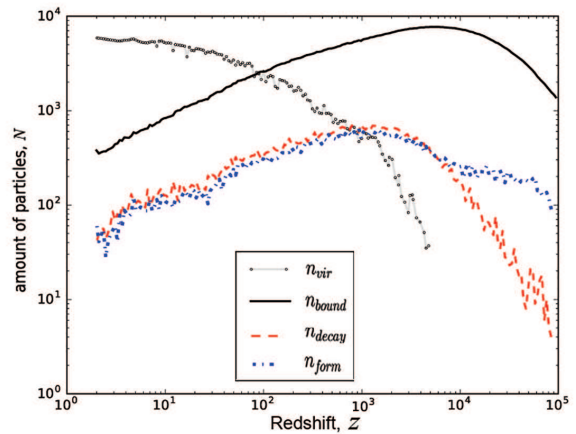}
    \caption {Number  of PBH particles $n$ as a function of redshift $z$ for a  PBH fraction $f_{PBH} = 1.0$, where $n$ represents various sub-sets of PBH particles (see legend): $n_{bound}$ -- a total amount of gravitationally bound particles in the simulation;  $n_{form}$ -- an amount of particles, which are currently (i.e. between two snapshots) forming a pair; $n_{decay}$ -- particles in a pair which is decaying at the given redshift $z$. And $n_{vir}$ represent particles, which are currently  part of a virialised structure (see Section \ref{sec:virial}).
    }
    \label{fig:form}
\end{figure}

After that, we evaluate the rates of classical bound pairs formation and pair disruption, depending on redshift $z$ (see Figure \ref{fig:form} for $f_{PBH} = 1.0$).

As one can see from this Figure, the rates of formation and disruption of pairs both grow rapidly towards the radiation-matter equipartition epoch, and then, shortly after, start to decay due to the process of matter virialisation (which we discuss in Section \ref{sec:virial}).

Before  going any  further,   we have to  check  that this behaviour is  not caused by   artefacts of the numerical calculation. Therefore,  in what follows we study the problem of pair stability as a function of the  softening length,  time integration step and multi-particle interactions, including the  formation of virialised clusters.

%%%%%%%%%%%%%%%%%%%%%
%%%%%%%%%%%%%%%%%%%%%
\subsection{Softening Length}          \label{sec:soft}
\begin{figure}
    \centering
    \includegraphics[width=0.47\textwidth]{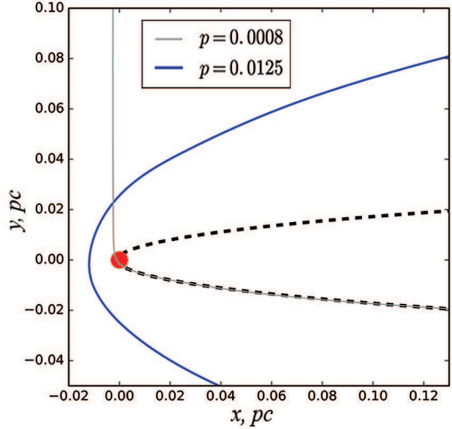}
   \caption{Trajectories of particles with various pericentre distances.  Blue solid line correspond to the simulated  orbit with  $p = 0.0125$ pc   and  light gray correspond to $p = 0.0008$ pc . The  \textit{dashed black line} represents  the theoretical trajectory for the particle with $p = 0.0008$ pc. The  red  solid circle represents the softening length $\epsilon = 0.001$.
	}
    \label{fig:soft_traj}
\end{figure}
In most of the modern N-body codes the gravitational potential is \textit{softened} in order to avoid the singularity at small distances between particles, which may result in close-to-zero timesteps and unreasonably long calculation times. In \texttt{GADGET-2} the single particle density distribution function $\tilde{\delta}$ is the Dirac $\delta$-function convolved with a normalized gravitational softening kernel
of comoving scale $\epsilon$. In \texttt{GADGET-2}   the spline kernel \citep{mon-lat},  often used in SPH,   is chosen and set  to $\tilde{\delta} = W(|x|, 2.8\epsilon)$, where $W(r$) is given by
\begin{equation} \label{eq5}
W(r,h) = \frac{8}{\pi h^3} \begin{cases} 1-6(\frac{r}{h})^2+6(\frac{r}{h})^3, & 0 \leq \frac{r}{h} \leq \frac{1}{2}, \\
2(1-\frac{r}{h})^3, & \frac{1}{2} \leq \frac{r}{h} \leq 1, \\
0, & \frac{r}{h} > 1 \end{cases}
\end{equation}
(see \citet{gadget})

In the context of our simulations this means,  that if two gravitationally bound particles have a very small pericentre distance $p=a(1-e)$, they may get \textit{"artificially"} disrupted due to the gravitational potential being calculated as $-Gm/\epsilon$ at such distances, as  implied by \ref{eq5}.  This might lead to significant numerical errors in the calculated velocities of the particles. This effect  is shown in  Figure \ref{fig:soft_traj} for a particle orbit with $p = 0.0008$ pc, where  the gravitational softening $\epsilon = 0.001$ pc (exactly as in our main simulation set). On the other hand, a particle with an orbit with $p = 0.0125$ pc does not get disrupted.

In order to ensure  that our choice of softening length $\epsilon = 0.001$ pc is reasonable, we compare two simulations with different softening lengths $\epsilon_1 = 0.001$ pc and $\epsilon_2 = 0.02$ pc, while  we keep the PBH fraction $f_{PBH} = 1.0$ and the other initial parameters.  For both simulations we pick up the pairs which are getting disrupted during the \textit{given} snapshot and evaluate the mean anomaly\footnote{i.e.  the fraction of an elliptical orbit's period that has elapsed since the orbiting body passed its pericenter}
at the \textit{previous} snapshot for these pairs. This allows us to trace the moment of pericentre passage for every pair. If at the \textit{previous} snapshot we notice that the pair is about to pass the pericentre, but at the given snapshot it is  disrupted, and moreover,  if the pericentre distance for this pair is less than the softening length ($p < 2.8\epsilon$), then we conclude, that the disruption was caused by the \textit{softening}.
\begin{figure}
    \centering
    \includegraphics[width=0.47\textwidth]{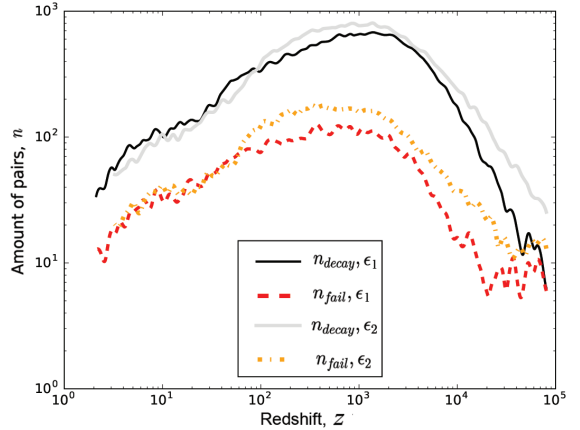}
   \caption{ \textit{Non-solid lines:} amount of pairs disrupted due to gravitational softening in both simulations (PBH fraction $f_{PBH} = 1.0$), \textit{Solid lines:} total amount of pairs disrupted per snapshot in both simulations. Where $\epsilon_1 = 0.001$ pc and $\epsilon_1 = 0.02$ pc are the softening lengths in each simulation, accordingly (see legend).
	}
    \label{fig:soft_fail}
\end{figure}

In  Figure \ref{fig:soft_fail} we display the respective amounts of pairs disrupted due \textit{softening} of the gravitational potential, and the total amount of disrupted pairs per snapshot for each simulation. As can be seen,  the amount of pairs disrupted due to \textit{softening} (or \textit{"artificially"}, as we noted above) is relatively small, compared to the total amount of disrupted pairs. In the case of simulation with  smaller softening length $\epsilon_1 = 0.001$ pc the amount of \textit{"artificially"} disrupted pairs is approximately 20\% less than in the case of the simulation with $\epsilon_2 = 0.02$ pc. However, it is worth noting, that this improvement comes at the cost of 10 - 15 times larger CPU-hours requirements.  Therefore, we may conclude that our choice of the gravitational softening length is reasonable and that a  further reduction is probably unnecessary.

%%%%%%%%%%%%%%%%%%%%%
%%%%%%%%%%%%%%%%%%%%%
\subsection{Time integration numerical  errors} \label{sec:num_err}
\begin{figure}
    \centering
    \includegraphics[width=0.47\textwidth]{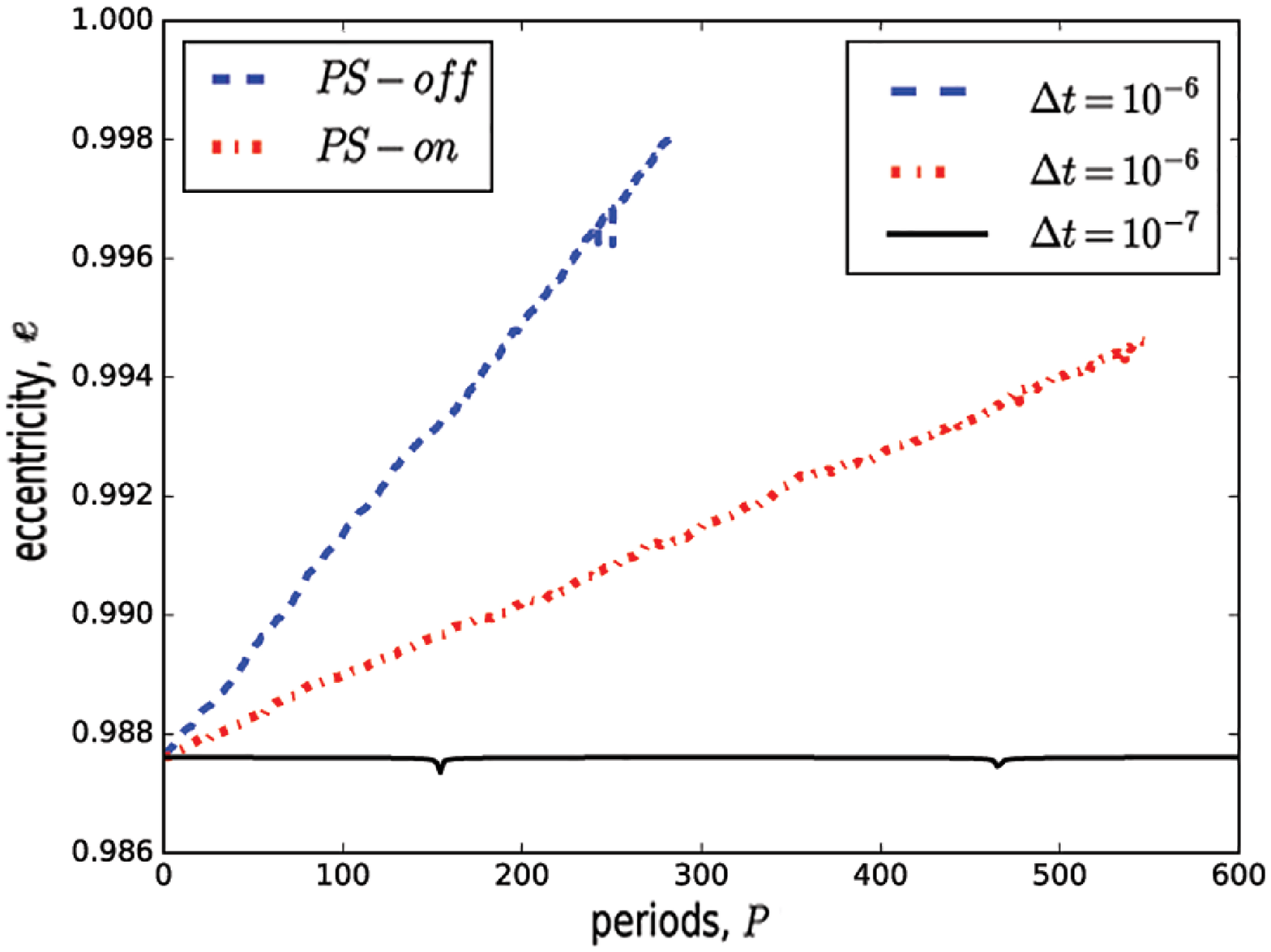}
    \includegraphics[width=0.47\textwidth]{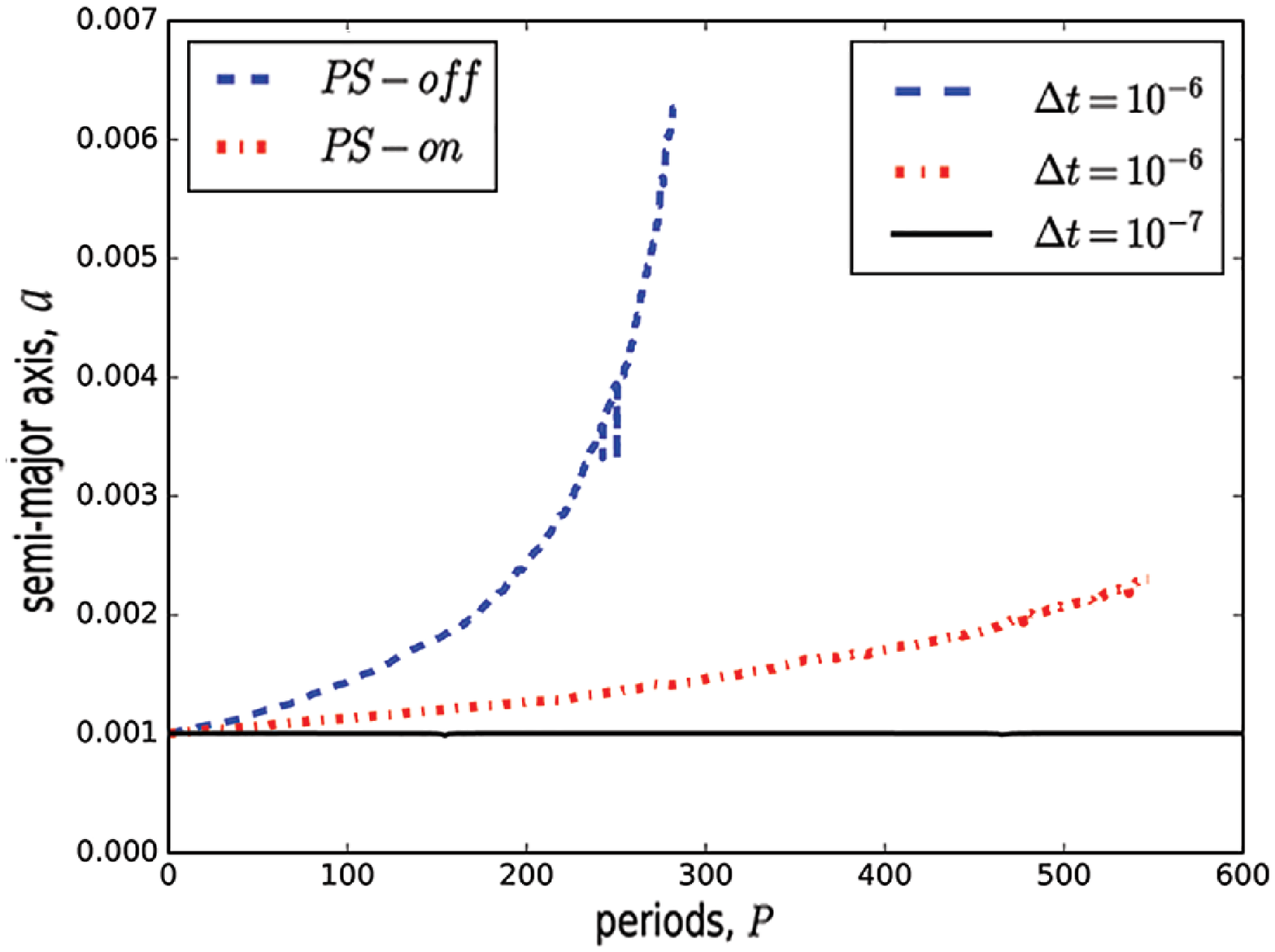}
    \caption{ Analysis of the orbital change of a two body system dues to numerical time integration errors in GADGET-2.  The   binary pair eccentricity $e$  (top panel) and the semi-major axis $a$ (\textit{bottom panel}) as a function of time in units of  the orbital periods.
The    \textit{Blue dashed line} corresponds to the simulation with the  \texttt{PSEUDOSYMMETRIC} property on and timestep $\Delta t = 10^{-6}$, \textit{red dot-dashed line} represents the simulation with \texttt{PSEUDOSYMMETRIC} property off and timestep $\Delta t = 10^{-6}$. \textit{Solid black  line} represents the simulation with a timestep $\Delta t = 10^{-7}$. Results from this simulation are not affected by the \texttt{PSEUDOSYMMETRIC}  compile option}
    \label{fig:odecay}
\end{figure}
In \texttt{GADGET-2} the \textit{"leapfrog"} time integration scheme is use  which, compared  to ordinary numerical integration methods like a high order Runge-Kutta, provides a much better orbital energy conservation and is more stable overall. However, depending on the timestep, the integration errors might still be considerable for a specific problem, which is why we perform a test set of  ideal 2-body simulations using the \texttt{GADGET-2} code. The initial parameters of these simulations are almost exactly the same as for our main simulation set, except that  we made an additional simulation with a smaller timestep $\Delta t = 10^{-7}$ (compared to  $\Delta t = 10^{-6}$ used in our main set). We also  turned off the box periodicity and eliminated the Hubble expansion, switching from comoving coordinates to physical ones.
Precisely two particles are placed into an otherwise empty cube with initial conditions for coordinates and velocities corresponding to the pair's eccentricity $e = 0.9877$  and the orbital period $T = 0.0123$ Gyr\footnote{1 Gyr = $10^9$ years}.
We have run the simulation for a  total time $t \approx 1000 \times T = 12.5$ Gyr.
As it can be seen from Figure \ref{fig:odecay},  for the case with a smaller timestep $\Delta t = 10^{-7}$ (solid black line) the  orbit remains surprisingly stable over the course of the whole simulation. However, for certain simulations in our main set (depending on the PBH fraction $f_{PBH}$ and the total amount of particles $N_{total}$) such a small timestep value may result in unreasonably long calculation times.

In order to reduce the impact of the numerical errors with higher timestep values we enable the compile flag \texttt{PSEUDOSYMMETRIC} in  the \texttt{GADGET-2} Makefile.
When this option is set, the code will try to "anticipate" timestep changes by extrapolating the change of the acceleration into the future.
This can in certain idealised cases (such as ours) improve the long-term integration behaviour of periodic orbits \citep{gadget}.
As one can see from Figure \ref{fig:odecay}, with \texttt{PSEUDOSYMMETRIC} option enabled, the orbital parameters of the pair decay much slowly  compared to the case when this option is disabled. In less than 300 periods the value of the semi-major axis for the simulation  without \texttt{PSEUDOSYMMETRIC} flag increases by more than $\sim$500\%, compared to only $\sim$40\% for the case with \texttt{PSEUDOSYMMETRIC} flag enabled.
\begin{figure}
    \centering
    \includegraphics[width=0.47\textwidth]{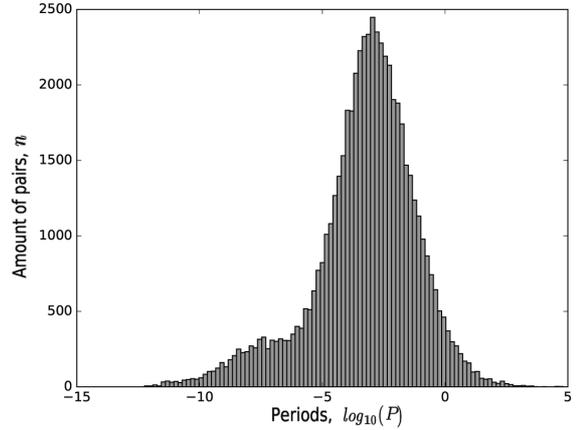}
   \caption{ The distribution of pair disrupting events throughout the whole simulation as a function ($f_{PBH} = 1.0$) of  $\log_{10}(P)$, with $P$ being  the number  of orbital  periods passed before the pair is   disrupted. I.e., "0" represents one full period.
	}
    \label{fig:periods}
\end{figure} 
However, in the main simulation set,  less than 5 pairs per snapshot have survived for even 100 periods before their subsequent disruption, while the absolute majority of pairs tend to disrupt after less than 1 period  (see Figure \ref{fig:periods}).
Thus, we conclude that disruption happens due to other reasons rather  than numerical errors. 

%%%%%%%%%%%%%%%%%%%%%
%%%%%%%%%%%%%%%%%%%%%
\subsection{Virialisation of Matter}   \label{sec:virial}
According to  most models of PBH binaries, the mechanism responsible for  pair breaking is tidal disruption by nearest neighbours (e.g, see \citet{yacine, raidal}).
In our  unperturbed $\Lambda$CDM model with Poissonian initial conditions, virialised structures containing many particles start to emerge at redshifts $z \approx 10^3$. On the other hand, in classical WIMP DM simulations structures start to form much later, at $z<100$ \citep{cosmogrid}, and in general, it is worth noting, that normal dark matter haloes would differ significantly from PBH dark matter haloes. In this work we \textit{do not} study the behaviour of PBH pairs in haloes consisting of \textit{normal DM}, as it was mentioned above (see Section \ref{sec:simulations}).
As the density inside a newly formed DM halo starts to rise, the distance between particles decreases, naturally, the distance to the nearest neighbour also decreases.

\begin{figure}
    \centering
    \includegraphics[width=0.47\textwidth]{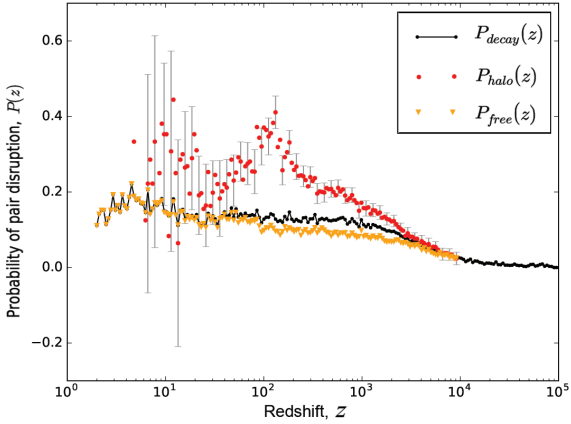}
   \caption{ Probabilities of pair disruption: inside PBH DM halos $P_{halo}$  (\textit{red dots}), outside halos $P_{free}$ (\textit{orange triangles}) and the total probability of pair disruption per snapshot $P_{decay}$ (\textit{black solid line}) as a function of redshift $z$. These results correspond to  the simulation with  $f_{PBH} = 1.0$
	}
    \label{fig:prob}
\end{figure}

As shown in  Figure \ref{fig:form}, pairs formation and  decay rates start to be damped at  $\approx 10^3$, while the rate of orbit destabilisation  rapidly  increases. In order to identify virialised structures formed by PBH particles, we analysed our simulations with the publicly available halo-finding tool \texttt{AHF}\footnote{http://popia.ft.uam.es/AHF/},
considering a minimum amount of PBH particles per  halo $N_{min} = 10$ and an  overdensity parameter $D_{vir} = 200$.

In Figure \ref{fig:form}, the  \textit{grey connected dots} represent the amount of particles, which are  currently  part of a  PBH-matter halo.
Tracking the indices of the particles with the  \texttt{AHF} tools  allows us to classify all gravitationally \textit{bound} particles $n_{bound}$ into two sub-sets: particles that are bound inside  some of the  PBH  halo $n_{halo}$, and  particles that are gravitationally bound to each other, but do not belong to any halo, $n_{free}$. Obviously, $n_{halo}+n_{free}=n_{bound}$. Since we know the indices of all the gravitationally bound particles, as well  those for the  disrupted particles,  we can now  obtain the probabilities of pair disruption inside the PBH halo $P_{halo}(i)$, outside the halo $P_{free}(i)$, and the total probability $P_{decay}(i)$ of pair disruption at the a given snapshot $i$ as follows:  
\begin{equation}
P_{halo}(i) = \frac{m_{halo}(i)}{n_{halo}(i-1)},
\end{equation}
\begin{equation}
P_{free}(i)= \frac{m_{free}(i)}{n_{free}(i-1)},
\end{equation}
\begin{equation}
P_{decay}(i)= \frac{n_{decay}(i)}{n_{bound}(i-1)},
\end{equation}
where $m_{halo}(i)$ and $m_{free}(i)$ represent the amount of disrupting particles inside and outside of the PBH  halo respectively at the $i$-th snapshot, and $m_{halo}(i) + m_{free}(i) = n_{decay}(i)$, while $n_{halo}(i-1) + n_{free}(i-1) = n_{bound}(i-1)$ represent the amount of bound particles at $(i-1)$-th snapshot.

As one can see from Figure \ref{fig:prob}, the probability of pair disruption inside PBH  halos is significantly higher compared to that one  outside halos,  as soon as the \textit{"first"} virialised structures begin to emerge between redshifts $z=10^4$ and $z=10^3$, which corresponds to radiation-matter equipartition, or even prior that. Large error values at lower redshifts ($z<10$) are related to extremely small amount of pairs still remaining in halos.

The \textit{"first"} gravitationally-bound structures that we are able to find with \texttt{AHF} halo finder appear to be small clusters of the size of $\sim$10 particles (which is the value we used in AHF to define the minimum  number of particles per halo  $N_{min} = 10$). Such tiny  structures can hardly be considered halos in the traditional meaning. In the case of our simulation set, "classic" halos with many particles can be detected only in simulations with larger PBH fraction, such as $f_{PBH}=1.0$ and $f_{PBH}=0.1$. In reality, however, gravitationally-bound structures as small as 3-10 particles do exist in all of our 5+1 simulations.
Finding such gravitationally-bound structures at high redshifts, in simulations with low values of PBH fraction, using  with conventional methods,  represents a significant challenge.
This seem to happen due to conceptual restrictions of the cosmological halo-finding techniques based on Bound Density Maxima algorithms, which require a measurable level of density contrast for a \textit{reliable} halo detection, while at earlier redshifts the density contrast caused by such a small clusters is virtually non-existent.
Attempting to use halo finders based on Friends-Of-Friends (\textit{FOF}) algorithm (see \citet{fof}) does not yield any meaningful results, since for our case the threshold distance can not be chosen unambiguously, as the variations of inter-particle distance both in the field and inside the clusters are large. I.e., the result will depend on the choice of threshold distance.

We conclude that the fraction of virialised matter is actually higher at high redshift, $z\geq 10^3$, than is shown in our Figure \ref{fig:form}, so we cannot reliably estimate the impact of clusters of PBH particles on pair disruption at that time. At later times, $z\sim 10^2$, when virialised PBH matter is condensed in large halos, our estimates of virialisation impact should be correct.

%%%%%%%%%%%%%%%%%%%%%
%%%%%%%%%%%%%%%%%%%%%
\section{PBH mergers and $\lowercase{f}_{PBH}<1.0$}  \label{sec:merger}
In the previous Section we came to the  conclusion  that numerical errors do not seem to play any significant role  in the process of pair formation and disruption.  Instead,  it appears that the  most significant factor in that process is matter virialisation,  that starts happening at  $z \approx 3000$. This qualitative result seems to be in good accordance with estimates recently obtained by \citet{clustering1} and \citet{clustering2}.

\begin{figure}
    \centering
    \includegraphics[width=0.47\textwidth]{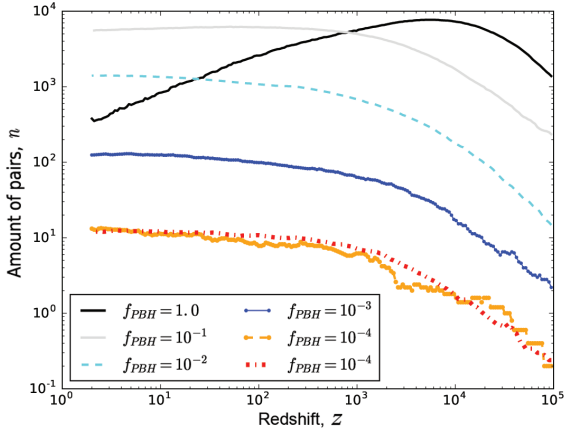}    
    \caption{Number of  PBH pairs   that remain \textit{gravitationally-bound} during the entire simulation as a function of redshift.  The total number of particles is fixed to $N=10^4$.  Simulations with   $N>10^4$ are rescaled to that value.  The different lines correspond to simulations with different   fractions of dark matter in the form of PBHs,  varying from  $f_{PBH} = 1.0$ to $f_{PBH} = 10^{-4}$ (see legend for specific symbols).   The \textit{orange dashed line} represents the simulation with $f_{PBH} = 10^{-4}$ and the total number of particles $N=10^5$, while \textit{red dot-dashed} line represents the simulation with the same PBH fraction but for   $N=10^6$.
    }
    \label{fig:frac}
\end{figure}

Now we can proceed to our main goal: to  estimate  the merger rate of PBH pairs from our simulations with different  fraction of dark matter in the form of PBHs. The number of gravitationally-bound particles for each simulation is shown in Figure \ref{fig:frac}. Interestingly enough, the number of binaries scales roughly proportional to $f_{PBH}$, which  resembles the result obtained by \citet{kavanagh}.
For  each \textit{given snapshot} we look for PBH pairs that were formed at the \textit{previous snapshot} and survived (see Section \ref{sec:analysis}). We consider such a pair as a \textit{newly formed} one.  Then we calculate the mean anomaly of the pair and observe it gradually changing at  each snapshot, during which the pair remains gravitationally bound. If particles  pass  the pericentre, we calculate the left and right terms  of the merge condition (see Eq. \ref{eq1}) at the given snapshot. If the pair is disrupted after passing the pericentre (see Section \ref{sec:soft}) we calculate these two terms of the  merger condition  using the orbital parameters from the previous snapshot. 
In this way we do not miss those  pairs which were "unfairly" disrupted when their particles passed each other within a distance shorter than the softening length. When  this  condition is satisfied, we declare the pair  as \textit{"merged"}. It is also worth noting that for  PBH of  masses   $M=30 M_\odot$,  the  maximum possible pericentre for the merging pair is $2.6\times 10^{-4}$ pc, which can be calculated assuming the merger time for the pair to be  less than $\sim 10$ Gyr.
In practice, however, after merging the particles still remain where they are  and the pair, in fact, does not get replaced by  twice a heavier particle. Since  the amount of pairs should decrease after merging, in order to avoid over-abundance of acting particles in the simulation we artificially "delete" one of the particles from the merged pair. So that when such pair is disrupted, only one of the particles is now considered as "free", and another one is ignored for the rest of the simulation.
We have also checked that our results on the merger rate do not change significantly if we keep both particles of a merged pair, or if we delete both of them.

\begin{figure}
    \centering
    \includegraphics[width=0.47\textwidth]{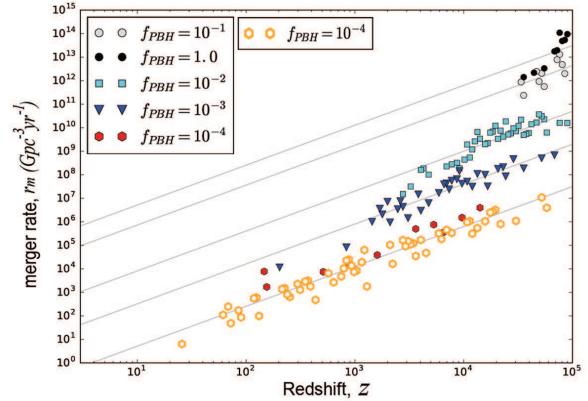}
    \caption{The merger rate of PBH binaries, $r_m$, as a function of redshift $z$ (see legend for specific symbols). \textit{Red filled hexagons} represent the results for the  simulation with $f_{PBH} = 10^{-4}$ and $N = 10^5$  particles, while \textit{orange-bordered white hexagons} represent the high resolution simulation with $f_{PBH} = 10^{-4}$ and $N = 10^6$ particles.  The extrapolated values to the corresponding points for different PBH fractions are represented by \textit{light-gray straight lines}}
    \label{fig:mergers}
\end{figure}

In  Figure  \ref{fig:mergers} we show the corresponding merger rates, $r_m$ for each of our simulations listed in Table 1. As can bee seen  in that  Figure,  we fail to detect merger events below  a   certain redshift,  for all the  five  simulations. Therefore, we decided to make an additional simulation (described in Section \ref{sec:simulations}) with an increased number of  PBH particles $N = 10^6$ in order to verify how  the particle number  would improve this situation. The \textit{orange-bordered white hexagons} in  Figure \ref{fig:mergers}
represent this high resolution  simulation,  with 10 times more particles (compared to the main  simulation set with $f_{PBH} = 10^{-4}$ and $N = 10^5$  particles) one can observe significantly more merger events, while the merger rate per Gpc$^3$ per year for the PBH fraction remains approximately the same.
Additionally, as we mentioned in \S \ref{sec:intro}, we also considered mergers caused by GW radiation in close hyperbolic encounters, and it appeared that only 0.77\% of all mergers throughout all our 6 simulations had hyperbolic orbits, and only the simulations with the higher fraction of PBH (1.0 and 0.1) had just a measurable number of such mergers (see Table \ref{tab:GR}).

In order to register merger events at $z=3$ for a PBH fraction $f_{PBH} = 10^{-3}$ we would have to increase the amount of particles in the simulation from $N = 10^5$ to approximately $N = 10^8$, which would  increase the computational time by an order of 100 to 1000, resulting in a simulation of $>10^6-10^7$ of CPU-hours. As an alternative, we can just simply extrapolate our  result to the given redshift. The results of this extrapolation are shown in Figure \ref{fig:mergers} as \textit{light-gray straight lines}. 
In the case with $f_{PBH} = 10^{-3}$ , the merger rate roughly corresponds to $r_m \approx 10^2$ $Gpc^{-3} yr^{-1}$ at redshift $z \approx 3$. However, despite the fact that  the result for this PBH fraction matches the estimation of the merger rate by the LIGO collaboration \citep{ligo7} and by other groups \citep{yacine, raidal, clustering2}, such an extrapolation might not be well suited for the estimation of the merger rates for  PBH fractions  $f_{PBH}>0.001$ and it merely represents a qualitative result here. As we discussed in Section \ref{sec:virial}, the formation of large gravitationally-bound structures (such as DM halos) might prevent merger events from occurring altogether, while on the other hand, a choice of a more realistic model (for example, the one with the proper account of GR effects in  dense clusters, as discussed in a recent paper by  \citet{trashorras}) may complicate things even further.

%%%%%%%%%%%%%%%%%%%%%
%%%%%%%%%%%%%%%%%%%%%
\section{Conclusions}   \label{sec:conclusions}
The main purpose of this work is to  expand our  understanding of the process of formation, disruption and merging of bound and unbound pairs of primordial black holes (PBHs) as one of the possible candidates  responsible for the  events registered by The Advanced Laser Interferometer Gravitational-Wave Observatory (LIGO), including
\citet{ligo1, ligo2, ligo7, ligo3, ligo5, ligo4, ligo8}.

We performed  a series of cosmological simulations in a box with periodic boundary conditions, where either all of the dark matter or  a variable  fraction $f_{PBH}$ is represented by primordial black holes (PBHs) with  30 M$_\odot$ masses. We assume that "normal" dark matter constitute an homogeneous fluid  with no density fluctuations, that only contributes to the evaluation of Hubble  function. We showed  that this assumption is reasonable for  $z\ge 10$.  However, we have pushed our simulations down in redshift, up to $z=3$, at the cost of accuracy. Nevertheless, errors associated  to  box periodicity would not allow us to run the simulations down to  redshift $z = 0.0$.

In order to account for the radiation dominated stage of the Universe expansion we use a modified version of public N-body code \texttt{GADGET-2} \citep{gadget} with specific parameters tuned  to calculate the orbital parameters of the forming PBH pairs with  sufficient precision  to accurately track their orbits  until they merge or get disrupted by interactions with  their neighbours. Additionally, we concluded that the number of binary captures during hyperbolic close encounters due to GR effects is  negligible compared to the number of pure Newtonian captures.

We devote a significant amount of effort in order to verify  that the measured pair disruption  is not caused by the intrinsic numerical errors associated to any N-body computation  of  gravitational  orbits  (see Figure \ref{fig:form} and Section \ref{sec:num_err}).
As a result, we ensured, that this is not the case, and that the main factor responsible for the processes of PBH pairs formation, disruption and merging is the emerging of gravitationally bound structures around redshifts of $z \approx 3000$, which roughly corresponds to the radiation-matter equipartition epoch.

We have also reported  our findings on the merger rates of PBH pairs, including hyperbolic mergers, which represent 0.77\% of the total number of mergers across all 6 simulations. By comparing the merger rates of simulations with the same PBH fraction $f_{PBH} = 10^{-4}$ and varying amount of particles, $N = 10^5$ and $N = 10^6$,  we demonstrated  that the sample size affects the amount of registered merger events but, at  the same time,  the merger rate  remains approximately the same. Then we extrapolate the  merger rate trends for all PBH fractions up to the redshift $z=3$, where we estimate a merger rate value of $r_m \approx 10^2$ $Gpc^{-3} yr^{-1}$ for PBH fraction of $f_{PBH} = 10^{-3}$,  which nicely matches  the merger rate estimated  by the  LIGO collaboration \citep{ligo7}
and  other groups  \citep{yacine, raidal}.

However, our estimations of the merger rates for  larger values of PBH fractions ($f_{PBH} > 0.01$) might be significantly overestimated, because the  simple extrapolation does not account for the influence of the processes related to  matter virialisation.
On the other hand, as it has been recently shown by \citet{luca_accretion}), accretion processes existing prior to redshift $z \sim 10$ might also affect the PBH abundance. As discussed by these authors, the accretion rate depends significantly on the relative velocity between the PBHs and baryonic matter.  Therefore, virialisation might play even a more important role in the evolution of PBH pairs  than we have  suggested in Sec. \ref{sec:virial}.

In this paper we have discussed the qualitative implementations of this phenomena -- but, obviously, a more thorough quantitative analysis is required. Therefore, in a future work we plan to create new numerical simulations, where the \textit{"normal"} DM would be represented as a separate type of N-body particles with their corresponding initial conditions compatible with the $\Lambda CDM$ cosmological model. 

%%%%%%%%%%%%%%%%%%%%%
%%%%%%%%%%%%%%%%%%%%%
\section*{Acknowledgements}
We are eternally indebted to P. B. Ivanov for many illuminating discussions and comments. The work of MVT and SVP was supported by RFBR grant 19-02-00199.
GY acknowledges financial support from {\em Ministerio de Ciencia, Innovaci\'on y Universidades / Fondo  Europeo de DEsarrollo Regional,}  under research  grant  PGC2018-094975-C21.
We would like to thank the {\em Red Espa\~nola de Supercomputaci\'on} for granting us access to the {\sc MareNostrum } supercomputer at the BSC-CNS where the simulations used in this work have been run.

\section*{Data availability}
The data underlying this article will be shared on reasonable request to the corresponding author.

%%%%%%%%%%%%%%%%%%%%%%%%%%%%%%%%%%%%%%%%%%%%%%%%%%

%%%%%%%%%%%%%%%%%%%% REFERENCES %%%%%%%%%%%%%%%%%%

\bibliographystyle{mnras}
\bibliography{pbh} 

%%%%%%%%%%%%%%%%% APPENDICES %%%%%%%%%%%%%%%%%%%%%

%%%%%%%%%%%%%%%%%%%%%%%%%%%%%%%%%%%%%%%%%%%%%%%%%

% Don't change these lines
\bsp	% typesetting comment
\label{lastpage}
\end{document}